\documentclass[10pt,english,roman,amsmath,amssymb,onecolumn,showpacs,casabs]{article}
\usepackage{amsmath,amstext,amssymb,graphicx,geometry,color,parskip,times,babel,setspace,subfig,
caption,floatrow,nccmath,titlesec,lipsum}

\title{\normalsize \textbf{Allee dynamics: Growth, extinction and range expansion}}
\author
{\footnotesize Indrani Bose\\
        \footnotesize {\it Department of Physics, Bose Institute}\footnotesize \\
       \footnotesize {\it 93/1, Acharya Prafulla Chandra Road, Kolkata-700009, India}\footnotesize\\
       \footnotesize {\it indrani@jcbose.ac.in}\footnotesize\\
       \footnotesize Mainak Pal\\
       \footnotesize {\it Department of Physics, Bose Institute}\footnotesize \\
       \footnotesize {\it 93/1, Acharya Prafulla Chandra Road, Kolkata-700009, India}\footnotesize\\
       \footnotesize {\it mainakpl3@gmail.com}\footnotesize\\
        \footnotesize Chiranjit Karmakar\\
       \footnotesize {\it Department of Physics, Indian Institute of Technology Guwahati},\footnotesize \\
       \footnotesize {\it Guahati-781039, Assam, India }\footnotesize \\
       \footnotesize {\it k.chiranjit@iitg.ernet.in}}
       
\date{}

\titlespacing*{\section}
{0pt}{1.5ex plus 1ex minus .2ex}{0.5ex plus .2ex}
\titlespacing*{\subsection}
{0pt}{1.5ex plus 1ex minus .2ex}{0.5ex plus .2ex}

\begin{document}
\maketitle

\begin{abstract}
In population biology, the Allee dynamics refer to negative growth rates below a critical 
population density. In this paper, we study a reaction-diffusion (RD) model of popoulation
growth and dispersion in one dimension, which incorporates the Allee effect in both the 
growth and mortatility rates. In the absence of diffusion, the bifurcation diagram displays 
regions of both finite population density and zero population density, i.e., extinction. The 
early signatures of the transition to extinction at the bifurcation point are computed in the 
presence of additive noise. For the full RD model, the existence of travelling wave solutions 
of the population density is demonstrated. The parameter regimes in which the travelling wave 
advances (range expansion) and retreats are identified. In the weak Allee regime, the 
transition from the pushed to the pulled wave is shown as a function of the mortality rate 
constant. The results obtained are in agreement with the recent experimental observations on 
budding yeast populations .
\end{abstract}



\maketitle

 \section*{\normalsize 1. Introduction}
 \label{intro}
 \normalsize
 \setlength{\baselineskip}{13pt}

 Biological systems are characterised by dynamics with both local and non-local components. In
 the case of well-mixed systems, one needs to consider only local dynamics based on
 reaction/growth kinetics. The concentrations of biomolecules like messenger RNAs (mRNAs) and proteins increase
 through synthesis and decrease through degradation. The density of a cell population is subjected to
 changes brought about by birth and death processes. In the latter case, depending upon specific conditions, the
 cell population acquires a finite density in the course of time or undergoes extinction. In the case of a spatially
 extended system, reaction-diffusion (RD) processes govern the dynamics of the system.
 The RD models have been extensively studied in the context of spatiotemporal pattern formation in a variety of
 systems \cite{cross,koch}. One possible consequence of RD processes is the generation of travelling waves which are
 characteristic of a large number of chemical and biological phenomena \cite{murray1,lewis}. The shape
 of a travelling wave is invariant as a function of time and the speed of propagation is a constant. Biological systems
 exhibit travelling waves of measurable quantities like biochemical concentration, mechanical deformation, electrical signal,
 population density etc. One advantage of travelling fronts in biological systems  is that for communication over
 macroscopic distances, the propagation time is much shorter than the time required in the case of purely diffusional
 processes. In this paper, we study cell population dynamics in a one dimensional (1d) spatially extended system described
 by the RD equation with the general form
 \begin{equation}
 \frac{\partial u(x,t)}{\partial t} = D \frac{\partial^2 u}{\partial x^2} + F(u)
 \end{equation}
where {\it u}({\it x}, {\it t}) denotes the population density at
the spatial location {\it x} and time {\it t}, D is the diffusion
coefficient and the term F({\it u}) represents the local growth rate
incorporating the Allee effect \cite{lewis,cour,kramer} implying
reduced per capita growth rates at low population densities. In the
case of the strong Allee effect, the growth rates becomes negative
below a critical population density resulting in population
extinction. The specific form of F({\it u}), proposed in an earlier
study \cite{beree}, is given in Eq.(10). The focus of the study was
to investigate the outcome of the combined processes of the Allee
effect and the collective movement of the population through
diffusion between the patches of a fragmented habitat. In our study,
the continuous space RD model is investigated for the first time and
we demonstrate the utility of the model in addressing issues like
population growth, extinction and range expansion {\it
vis-\`{a}-vis} some recent experimental observations on laboratory
populations of budding yeast \cite{dai1,dai2,sen,gandhi}.

\section*{\normalsize 2. Models of Growth and Extinction}
 \label{model}
 \normalsize
 \setlength{\baselineskip}{13pt}
 
 In the absence of diffusion, only the the local dynamics of the population are relevant and the major issue of
 interest is the  survival or extinction of the population in the steady state. In the case of logistic growth \cite{murray1},
 \begin{equation}
  F(u) = \rho u (1-\frac{u}{K})
 \end{equation}
 where $\rho$ denotes the exponential growth rate and K the carrying capacity (the population growth stops when {\it u} = K).
 The growth rate F({\it u}) satisfies the properties
 \begin{equation}
  F(0) = F(K) = 0
 \end{equation}
with the steady state {\it u} = 0 being unstable and the steady
state {\it u} = K being stable, a case of monostability.\newline
Also,
\begin{equation}
 F(u) > 0 \:for\: 0 < u < K, \:\:\:\: F(u) < 0 \:for\: u > K
\end{equation}
\begin{equation}
 F'(0) = \rho > 0, F'(u) < \rho \:for\: u > 0
\end{equation}
where the prime represents the derivative with respect to {\it u}.
The per capita growth rate, {\it f}({\it u}) = $\frac{F(u)}{u}$, is
a monotonically decreasing function of the population density, i.e.,
the per capita growth rate has a negative dependence on the
population density. One common departure from the logistic growth,
observed in natural and laboratory populations, is designated as the
Allee effect \cite{lewis,cour,kramer} with positive dependence of
the per capita growth rate on the population density till the
maximal population density is achieved. In the case of the strong
Allee effect, the local growth rate F({\it u}) is negative when the
population density {\it u} is less than a critical density
 $\theta$ known as the Allee threshold. The negative growth rate results in population extinction in the long time limit.
 The most well-studied functional form of F({\it u}), which illustrates the Allee effect, is
 \begin{equation}
  F(u) = \rho \: u\: (u - \theta ) \: (1-\frac{u}{K})
 \end{equation}
One now has
\begin{equation}
 F(0) = F(\theta) = F(K) = 0
\end{equation}
with the steady states {\it u} = 0 and {\it u} = K being stable, a
case of bistability, and the state {\it u} = $\theta$ being
unstable. The other conditions on  F({\it u}) are

\begin{equation}
 F(u) < 0 \: \:for \:\: 0 < u < \theta \:\: and \:\: u > K
\end{equation}

\begin{equation}
 F(u) > 0 \:\: for \:\: \theta < u < K
\end{equation}

The per capita growth rate {\it f}({\it u}) is negative below the
Allee threshold and reaches a maximum value at an intermediate
density. In the case of the weak Allee effect, the per capita growth
rate is always positive signifying the absence of an Allee
threshold. The maximal value again occurs at the intermediate
density. A possible origin of the Allee effect lies in the reduced
cooperativity amongst the individuals of the population al low
population densities \cite{cour,kramer,kor}. Examples include mate
shortages in sexually reproducing species, less efficient feeding
and reduced effectiveness of vigilance and antipredator defences \cite{cour,kramer}. Allee effects have been demonstrated in all
major taxonomic groups of animals, in plants\cite{kramer} as well
as microbial populations \cite{dai1,kaul}.

\begin{figure}
 \includegraphics[scale=0.5]{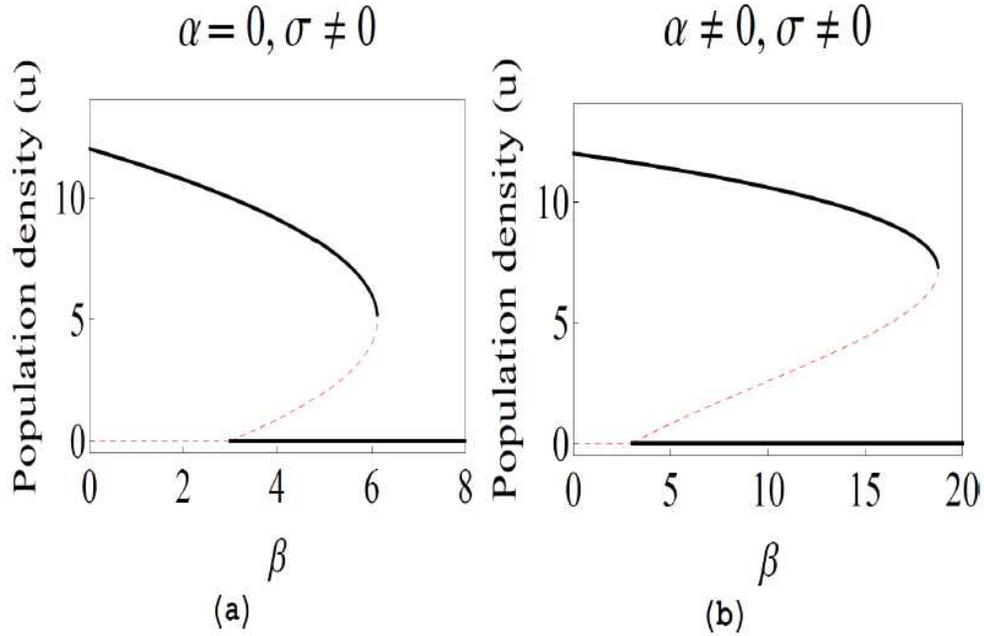}
 \caption*{Fig. 1 Steady state population density {\it u} versus the mortality rate constant $\beta$ for the cases
 (a) $\alpha$ = 0 and (b) $\alpha$ = 1.0 The other parameter values are $\rho$ = 3.0, K =12.0, $\sigma$ = 2.0.}
\label{fig1}
\end{figure}

There are several ways of modelling the Allee effects \cite{lewis,lewis1} with most of the models explicitly including the
Allee threshold as a parameter such that the per capita growth rate
is negative (positive) below (above) the threshold. A less
phenomenological model has been proposed by Berec {\it et al}. \cite{beree} which incorporates the Allee effect in both the birth
(reproduction) and death (mortality) processess. The function F({\it
u}) in this case is given by
\begin{equation}
 F(u) = (\rho + \alpha \: u) \:\: u \:\:( 1- \frac{u}{K} ) - \frac{\beta u}{1+\frac{u}{\sigma}}
\end{equation}
where $\rho$ is the basic birth rate, $\alpha$ is the Allee effect
coefficient with higher values of $\alpha$ signifying steeper
increases in the reproduction rate, $\beta$ denotes the mortality
rate constant when the Allee effect is present, i.e., the per capita
mortality rate increases (decreases) when the population density
{\it u} approaches zero (increases) and the parameter $\sigma$ is
the population density for which the mortality rate is halved. If
$\alpha$ = 0 ($\sigma$ = 0), there is no Allee effect on
reproduction (mortality). If both the parameters $\alpha$ and
$\sigma$ are zero, the model reduces to the usual logistic growth
model. In this paper, we study the RD model described in Eq. (1)
with the functional form of F({\it u}) as given in Eq. (10). Though
the model is of a general nature, our focus in this paper is on
microbial populations, motivated by some recent experiments on
laboratory populations of budding
yeast \cite{dai1,dai2,sen,gandhi}.

\section*{\normalsize 3. Allee Dynamics}
\label{allee}

We first report on the steady state properties of the
model in the absence of diffusion. Figures 1(a) and 1(b) exhibit the
steady state population density {\it u} versus the mortality rate
constant $\beta$ for the cases $\alpha$ = 0 and $\alpha$ = 1
respectively. In the region of bistability, the two stable state
branches, represented by solid lines, are separated by a branch of
unstable steady states, denoted by a dashed line. The upper stable
steady state represents a population with finite density whereas the
lower stable steady state ({\it u} = 0) corresaponds to population
extinction. The transitions between the bistable and monostable
regions occur via the saddle-node (fold) bifurcation at two
bifurcation points. In the region of bistability, if the initial
population density is above the dotted line, the population acquires
a finite density in the steady state. The population undergoes
extinction if the initial population density falls below the dotted
line. When $\alpha$ = 0 the steady states are given by
\begin{equation}
 u_{_{1}} = 0, u_{_{2}}, u_{_{3}} = \frac{1}{2}\Bigg[(K-\sigma) \pm \sqrt{(K+\sigma)^{2} - \frac{4K\sigma}{\rho}\beta }\:\Bigg]
\end{equation}
with {\it u}$_{_{1}}$, {\it u}$_{_{2}}$ being stable and {\it
u}$_{_{3}}$ unstable. The region of bistability is obtained in the
parameter regime

\begin{equation}
 \rho < \beta < \frac{\rho \:(K + \sigma)^{2}}{4 \:\sigma K}
\end{equation}
The boundary points of the regime represent the two bifurcation
points. When $\alpha$ $\neq$ 0 ( Fig. 1(b)), one obtains the steady
states through numerical computation. As depicted in Figs. 1(a) and
1(b) , a tipping point transition occurs at the upper bifurcation
point from a finite population density to population collapse.

\subsection*{\normalsize 3.1 {\it \textbf{Early signatures of regime shifts}}}
\label{early} 

Recently, a large number of studies have been carried
out on the early signatures of tipping point transitions involving
sudden regime shifts in systems as diverse as ecosystems, financial
markets, population biology, complex diseases, gene expression
dynamics and cell differentiation \cite{scheffer1,scheffer2,pal1,pal2}. The early signatures include
the critical slowing down and its associated effects, namely, rising
variance and the lag-1 autocorrelation function as the bifurcation
point is approached. Other signatures have also been proposed, e.g.,
an increase in the skewness of the steady state  probability
distribution as the system gets closer to the transition point \cite{scheffer1,scheffer2}. 
For our model system ($\alpha$ $\neq$ 0,
$\beta$ $\neq$ 0), we compute the variance and the autocorrelation
time (measures the time scale of correlation between the
fluctuations at different time points), through simulation of the
stochastic difference equation
\begin{equation}
 u(t+ \Delta t) = u(t) + F(u) \Delta t + \Gamma \:\xi \:\sqrt{\Delta t}
\end{equation}
where F({\it u}), with the form given in Eq. (10), is computed at
time {\it t} and the last term on the right hand side represents the
effect of stochasticity (noise). $\Gamma$ is the strength of the
additive noise and $\xi$ represents a Gaussian random variable with
zero mean and unit variance. Let $\delta${\it u}({\it t}) = {\it
u}({\it t}) - {\it u}$_{_{2}}$ represent the deviation of the
population density {\it u} at time {\it t} from the stable steady
state population density {\it u}$_{_{2}}$. The simulation data are
recorded after 1000 time steps (stationarity conditions reached) for
an ensemble of five hundred replicate populations. Fig. 2A shows the
plots of $\delta${\it u}({\it t} + $\Delta t$) versus $\delta${\it
u}({\it t}) for two distinct values of the bifurcation parameter
$\beta$. The first figure depicts the results far from the
bifurcation point, $\beta$ = 18.79. The parameter values are
$\vartriangle${\it t} = 0.01 and $\Gamma$ = 0.5 with the other
parameter values the same as in Fig. 1(b). Figures. 2B and 2C display
the autocorrelation time $\tau$ and the steady state sample variance
$\sigma^{2}$ as a function of the parameter $\beta$. The
autocorrelation time $\tau$ is computed as $\rho$ =
e$^{-\frac{1}{\tau}}$ where $\rho$ is the lag-1 autocorrelation
estimated by the sample Pearson's correlation coefficient \cite{dai1}.
\begin{equation}
 \rho = \frac{1}{n-1} \frac{\overset{n}{\underset{i=1}{\sum}\:\:} (u_{_{i}}(t) -
 \langle u_{t} \rangle) (u_{_{i}}(t + \Delta{t}) - \langle u_{_{t + \Delta{t}}} \rangle)}{S_{u_{_t}} S_{u_{_{_{t +\Delta t}}}}}
\end{equation}
where {\it i} denotes the sample index, {\it n} is the total number
of samples, $\langle u _{t} \rangle$ is the sample mean and
S$_{_u{_{_t}}}$ the sample standard deviation at time {\it t}. In
the computation, {\it n} = 500 and the time interval
$\vartriangle${\it t} is treated as the unit time interval.
\begin{figure}[t]
\begin{center}
\includegraphics[scale=0.523]{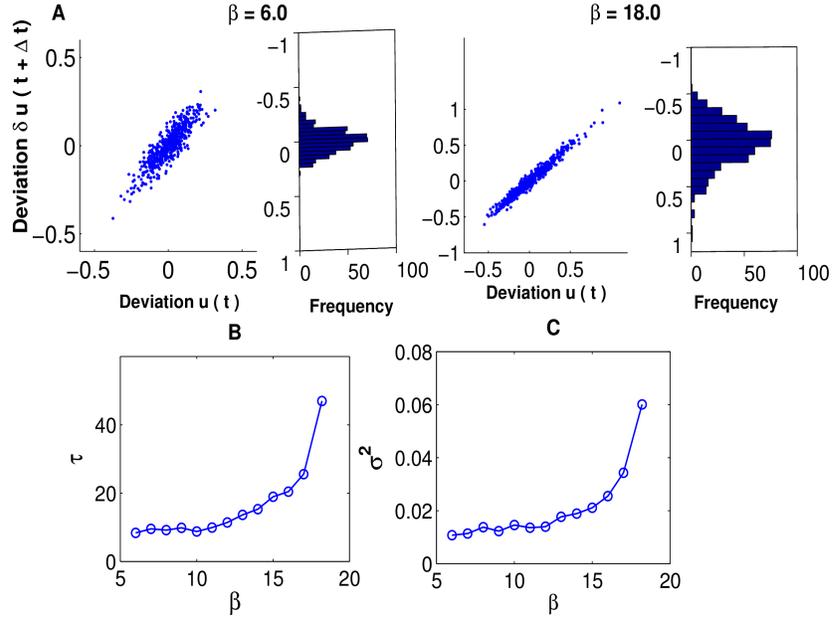}
\end{center}
 \caption*{Fig. 2  A. Temporal correlations and frequency distributions of deviations from the stable steady state at
 $\beta$ = 6.0 and $\beta$ = 18.0. B and C. Plots of autocorrelation time $\tau$ and steady-state variance $\sigma^{2}$ as
 a function of $\beta$.}
\label{fig2}
\end{figure}

Fig. 2 provides numerical evidence that both the variance and the
autocorrelation time increase as the bifurcation point is
approached. The return time equals the autocorrelation time \cite{dai1} 
and its increase in the vicinity of the bifurcation
point implies the critical slowing down. The results obtained in our
model study are in accordance with the experimental observations of
Dai {\it et al}. \cite{dai1} on laboratory populations of buding
yeast {\it S cerevisiae}. In the experiment, the Allee effect was
realized through the cooperative growth of budding yeast in sucrose.
Yeast cells are known to secrete an enzyme to hydrolyze
extracellular sucrose outside the cell thus creating a common pool
of the hydrolysis products, namely, glucose and fructose, which are
then transported into individual cells for utilization in growth
processes. The cooperative breakdown of sucrose results in the per
capita growth rate becoming maximal at the intermediate cell
densities when the competition for resources is not severe, a
feature of the Allee effect. Replicate yeast cultures with a wide
range of initial cell densities were subjected to daily dilutions
into fresh sucrose media. The dilution is equivalent to introducing
mortality in the population with higher dilution factors implying
larger mortality rates. An experimental bifurcation diagram
depicting the population density versus the dilution factor, which
serves as the bifurcation parameter, was mapped out. The bifurcation
was of the saddle-node type and a tipping point transition from a
population of finite density to population extinction was
identified. A simple two-phase growth model provided a good fit to
the experimental data. In the model, the population growth was slow
exponential at low densities and faster logistic at larger
densities, based on experimental evidence. The theoretically
predicted early warning signals of the tipping point transition were
also tested experimentally. The experimentally obtained bifurcation
diagram and the early warning signatures (Fig. 1(E) and Fig. 3 of
Ref. 8) are qualitatively similar to the theoretical plots (Figs. 1
and 2) computed in our model study. The model serves as a
microscopic model to describe population growth and extinction in
the budding yeast population and explicitly includes the parameters
describing the growth, the mortality and the Allee effect.

\section*{\normalsize 4. Reaction- Diffusion Model with Allee Dynamics}
\label{reaction}

We next consider the full RD model shown in Eq.(1). Depending on the form of F({\it u}),
two prototype cases are possible, monostability and bistability. The RD equation with
F({\it u}) given in Eq. (2) (logistic growth), was first considered by Fisher \cite{fisc} to
describe the spread of a favourable gene in a population. The equation describes monostability and is referred to as the
Fisher-Kolmogorov, Petrovskii and Piskunov (FKPP) equation. When F({\it u}) is of the form given in Eq. (6) and the conditions in
Eqs. (8) and (9) hold true, i.e., the strong Allee effect is considered, the situation is that of bistability. It has
been rigorously shown \cite{lewis,petro} that for suitable initial conditions {\it u}({\it x}, 0), the disturbance evolves into
a monotonic travelling front {\it u} = {\it q}({\it x} - {\it c}{\it t}), with constant speed {\it c}, joining the stable steady state with finite population
density to the steady state {\it u} = 0 (unstable in the case of monostability and stable in the case of bistability). Populations occupy new
territory through a combination of local growth and diffusion. If the population front moves from a region of finite {\it u} to one
with {\it u} = 0, the phenomenon is characterised as a range expansion or invasion. If the travelling front moves in the opposite direction, the
population retreats rather than advances.

In the case of the FKPP equation describing generalized logistic growth, a travelling wave solution with speed {\it c} satisfies
the relation {\it c} $\geq$ {\it c}$_{min}$ with
\begin{equation}
 c_{min} = 2\sqrt{\rho D}
\end{equation}

It has been shown that in the case of a compact initial condition,
the initial population distribution always develops into a
travelling population front with the minimum speed {\it c} = {\it
c}$_{min}$. The scaling {\it c} $\varpropto$ $\sqrt{\rho D}$ is
known as the Luther formula \cite{tyson} and holds true over many
orders of magnitude in many chemical and biological systems \cite{tayar}. 
A recent example pertains to the propagation of gene
expression fronts in a one-dimensional coupled system of artificial
cells \cite{tayar}. We next consider the case of reaction-diffusion
incorporating the Allee effect. When the strong Allee effect
conditions prevail, i.e., 0 $ < \theta <$ K in Eq. (6), the
travelling wave front has a unique speed given by \cite{murray1,lewis1,petro}

\begin{equation}
 v = \sqrt{\frac{D \rho}{2K}} (K - 2\theta)
\end{equation}

In conrtrast to the case of logistic growth in which the travelling front has a finite value for the minimum speed, the speed
{\it v} becomes zero at the so-called Maxwell point $\theta = \frac{K}{2}$  with reversal in the direction of motion of the
travelling wave as the Maxwell point is crossed. Also, the front has a unique speed instead of
a spectrum of possible values above or equal to a minimum speed. If $\theta$ and K are treated as  general parameters without any
specific physical interprtation, then the parameter regions 0 $< \theta <$ K and $-$ K  $ <\theta <$ 0 are associated with the strong and weak
Allee effects respectively. For $\theta$ $\leq$ $-$K, the Allee effect is absent and the growth rate is of the generalized logistic type.
Note that the parameter $\theta$ can no longer be treated as  critical density when it assumes negative values.

In the case of the FKPP equation, the minimal speed of the travelling front can be
written as {\it c}$_{min}$ = 2$\sqrt{F'(0) D}$. The travelling wave is designated as the pulled wave with the wave speed
depending solely on the growth rate at low population density and the diffusion
coefficient \cite{gandhi}. The expansion is dominated by the dynamics at low population density prevailing at the very edge of the
expanding wavefront which pulls the wave forward. The solution with the minimal speed defines the critical front while the
solutions for which the wave speed {\it c} is $>$ {\it c}$_{min}$ define the super-critical front. In the case of a pushed wave, the
minimal speed satisfies the relation
\begin{equation}
 c_{min} > 2 \sqrt{F'(0) D}
\end{equation}
The speed of the travelling front is now determined by the whole
front rather than by only the leading edge. In the monostable case
(weak Allee effect), the travelling front can be classified as
either pulled or pushed wave whereas in the case of bistability
(strong Allee effect), the travelling wave is always a pushed wave
with a unique speed greater than 2$\sqrt{F'(0) D}$. The theoretical
prediction of a pulled to pushed wave transition \cite{lewis1} has
recently been verified by Gandhi {\it et al.} \cite{gandhi} in a
laboratory population of budding yeast as the cooperativity in
growth was increased. The transition occurred at some intermediate
magnitude of the Allee effect in the regime corresponding to the
weak Allee effect.

The RD model studied in this letter with F({\it u}) as given in Eq. (10), has a more general form than the model most often studied in the
context of the Allee effect (F({\it u}) as given in Eq. (6)). The existence of a travelling wave solution in this model is
supported by the theorem proposed by Fife and McLeod \cite{fife}. The theorem states that the RD Eq. (1),
with F({\it u}$_{_{1}}$) =  F({\it u}$_{_{2}}$) = F({\it u}$_{_{3}}$) = 0 and
F$'$({\it u}$_{_{1}}$) $<$ 0, F$'$({\it u}$_{_{2}}$) $<$ 0,  F$'$({\it u}$_{_{3}}$) $>$ 0, has a travelling wave solution
{\it u} ({\it x}, {\it t}) = {\it U} ({\it x} $-$ {\it v}{\it t}) for exactly one value of the wave speed {\it v}.
The solutions {\it u}$_{_{1}}$ and {\it u}$_{{2}}$ describe stable steady states and the solution {\it u}$_{{_3}}$ an unstable
steady state. Also, U($-$ $\infty$) = {\it u}$_{{2}}$ and U($+$ $\infty$) = {\it u}$_{{1}}$, i.e., the travelling wave connects
 {\it u}$_{_{1}}$ and  {\it u}$_{_{2}}$. We assume that the parameter $\alpha$ = 0 in Eq. (10), i.e.,
 the Allee effect is present only in the mortality rate. In this case, the analytic solutions for the
 steady states are known ( Eq. (11)) and one can look for travelling wave solutions for the RD equation.
 In the following, we report on the results obtained from numerical computations.

 \subsection*{\normalsize 4.1 {\it \textbf{Range expansion, pushed and pulled waves}}}
\label{renge}

The 1d RD system has boundaries located at $-$L and $+$L
respectively. We solve the RD equation on Mathematica with the
compact initial conditions
\begin{equation}
 u (x, t) = u_{_{_2}} \:\: if \:\: x \leq 0
\end{equation}
\begin{equation}
 u (x, t) = u_{_{_1}} \:\: if \:\: x > 0
\end{equation}
The value of L is chosen to be 50, the diffusion coefficient D = 1
and the other parameter values are the same as in Fig. 1(a). Fig. 3
shows the computed travelling wave solutions at the time points {\it
t} = 20, 40, 60 and 80 and $\beta$ = 5. The front moves from the
left to the right with the population of finite density, {\it
u}$_{_{2}}$, advancing into the empty region, {\it u}$_{_{1}}$ = 0.
Fig. 4 shows the speed {\it v} of the travelling wave versus the
mortality rate parameter $\beta$ (solid circles). The speed was
computed by plotting the midpoint {\it x}$_{_{m}}$ of the density
profile versus time {\it t} and determining the slope of the
straight-line fit \cite{sen}. The larger solid black circles represent the speed
\begin{figure}[t]
 \includegraphics[scale=0.8]{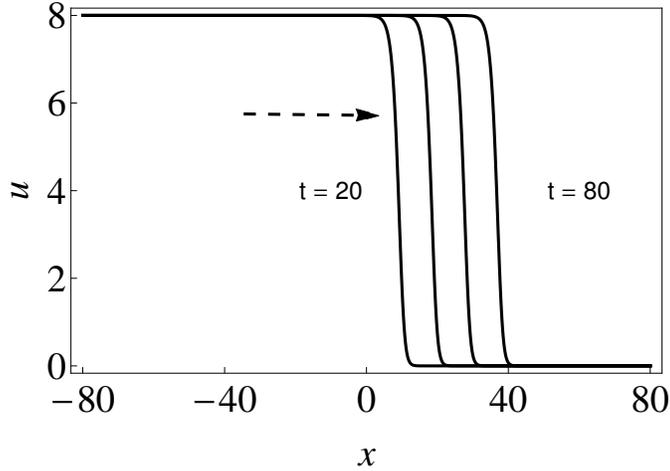}
 \caption*{Fig. 3  Travelling wave solutions of the RD Eq. (1) with F({\it u}) given in Eq. (10) for $\alpha$ = 0.
 The solutions are computed at the time points {\it t} = 20, 40, 60 and 80. The other parameter values are
 D = 1.0, $\rho$ = 3.0, K = 12.0, $\sigma$ = 2.0, and $\beta$ =5.0.}
\label{fig3}
\end{figure}
given by the formula, {\it v} = 2$\sqrt{F'(0) D}$ = 2$\sqrt{(\rho -
\beta) D}$, which corresponds to the speed of a pulled wave. The
computed speed of the travelling wave and that of the pulled wave
are nearly equal up to a transition point located within the big
circle. The RD equation reduces to the FKPP equation when the
parameter $\beta$ = 0. In the case of compact initial conditions, as
given in Eqs. (18) and (19), the exact expression for the travelling
wave speed is that of the pulled wave. The small difference with the
computed value can be ascribed to the finite size of the system and
the small but finite errors associated with the numerical
computation of speed. For $\beta$ $<$ $\rho$ = 3, the system is
monostable corresponding to the weak Allee effect regime. The strong
Allee effect regime is bistable with the parameter range as
specified in Eq. (12). At the transition point, $\beta$ $\approx$
1.823, the travelling wave changes its nature from the pulled to the pushed
wave. The speed of the travelling wave becomes zero at the Maxwell
point $\beta$ $_{MP}$ $\approx$ 5.6259. The travelling wave then
reverses its direction, i.e., the wave retreates from the empty
region rather than advance into it. The situation continues till the
upper bifurcation point $\beta = \frac{\rho(\sigma + K)^{2}}{4
\sigma K}$, marking the region of bistability, is reached. Close to
the Maxwell point, the travelling wave becomes very slow in a
markedly nonlinear fashion. The transition from the pulled to the pushed
wave occurs in the weak Allee effect regime, in agreement with the
experimental observation of the transition in a laboratory
population of budding yeast \cite{gandhi}. In the experiment, the 1d
RD system was set up by culturing the yeast populations in linear
arrays of wells on plates and through the exchange of small volumes
of the growth media between adjacent wells. In galactose or glucose
growth medium, the Allee effect is absent so that the RD process is
of the FKPP type with the travelling wave being a pulled wave with
the front speed obeying the Luther scaling (Eq. (15)). In sucrose
growth medium, the Allee dynamics come into play because the
cooperative breakdown of sucrose into glucose is required for
growth. The strength of the Allee effect was tuned by changing the
relative concentrations of glucose and sucrose in the growth medium.
The magnitude of the Allee effect was measured by the difference
between the maximal and the low density growth rates. The transition
from the pulled to the pushed wave was observed in the weak Allee
regime as the Allee effect was made progressively strong. In our
model study, the mortality rate constant $\beta$ was changed to
bring about the transition from the pulled to the pushed wave. Since
this parameter is related to the dilution factor in the original
yeast experiment of Dai {\it et al}. \cite{dai1}, our study suggests
that the transition in the nature of the travelling front could also
be tested by changing the dilution factor in the experimental RD
system. The strong Allee regime would then correspond to the range
of dilution factors in which the population density is bistable.

\section*{\normalsize 5. Conclusion}
\label{con}

In this paper, we have studied a RD model in 1d with the Allee
effect incorporated in the dynamics of the model. The model in its
full form has not been studied previously. The strength of the model
lies in the fact that the growth and the mortality rate constants
appear explicitly as parameters in the model which are
experimentally controllable. The model is general in nature and
applicable for the study of population dynamics based on growth and
diffusion. Though we could not derive analytical results for the
full model, the results obtained through numerical computations are
in agreement with
\begin{figure}[h]
 \includegraphics[scale=0.6]{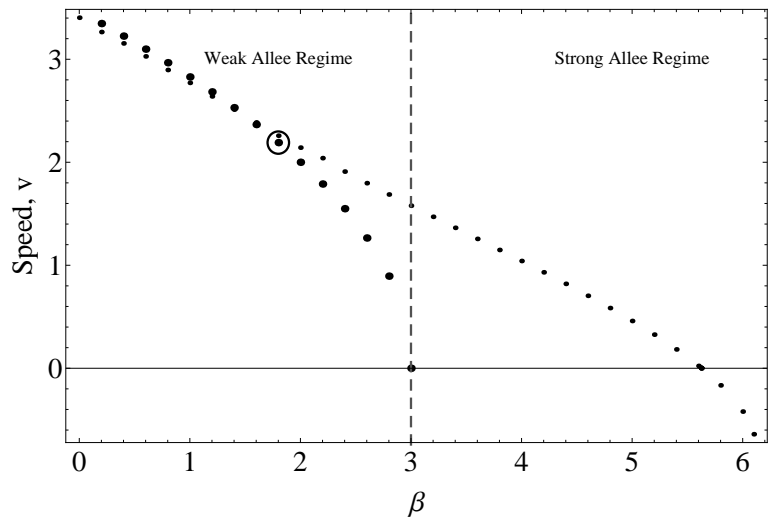}
 \caption*{Fig. 4 Speed {\it v} versus $\beta$. The big open circle denotes the transition region from the 
 pulled to the pushed wave. The larger solid black circles represent the speed of the pulled wave, 
 $ v = 2 \sqrt{(\rho - \beta)D}$}
\label{fig4}
\end{figure}
the experimental observations on budding yeast populations. The
model adds to the list of known models in population biology which
investigate issues like population growth, extinction and range
expansion. The knowledge of the early signatures of a tipping point
transition to extinction (Sec. 3.1) would be of use in the
eradication of harmful species like disease-causing bacteria,
viruses and pathogens. Recent advances in cancer research have drawn
parallels between tumour and population dynamics \cite{kor}. There
is some experimental evidence that the Allee effect forms an
important component in the dynamics of certain types of tumour. It
is possible that many tiny tumours are formed in different parts of
the body during the lifetime of an individual. Most of these tumours
become extinct because of their small size, signifying a negative
growth rate at low density as in the case of the Allee effect. It
has been found that xenograft transplantations of cancer cells into
mice with a functional immune system, are more likely to initiate
tumours when the number of injected cells is large. The basis of the
Allee effect may lie in cooperativity in the form of cancer cells
contributing to a common pool of diffusible growth factors required
for tumour proliferation. As emphasized in Ref. 12, the knowledge of
the proximity of a tumour close to a growth threshold would
facilitate in the alteration of drug doses. The detection of early
signatures similar to the ones discussed in Sec. 3.1 could
contribute in formulating therapeutic intervention protocols.

In the case of the full RD model with Allee-type dynamics, the
existence of a travelling wave solution indicates the possibilities
of both range expansion and retreat. The Allee efect makes it
possible, by controlling the parameter values, to prevent the range
expansion of harmful organisms or at least to slow down the speed of
the spread. The containment of the pulled and pushed invasions
requires different strategies: eradication of the invading
population at the very edge of the expansion in the first case and
eradication over the entire invasion front in the second case. The
RD model studied in the paper could be generalized to higher
dimensions, say, d = 2 and a stochastic component included in the
model. A recent study \cite{villa} indicates that in a 2d system,
discontinuous transitions are of the type shown in Fig. 1 could be
replaced by continuous transitions under the conditions of limited
diffusion, enhanced noise and quenched spatial heterogeneity.
Controlled experiments on laboratory populations of microbes may
offer the opportunity to test some of the theoretical predictions.

\subsection*{\normalsize Acknowledgments}

IB acknowledges the support by CSIR, India, vide sanction Lett. No. 21 (0956)/13-EMR-II dated 28.04.2014. 
CK acknowledges the support by  National Network for Mathematical and Computational Biology, India for carrying out part of the 
study. MP acknowledges support from Bose Institute, India for carrying out the research study.

 \end{document}